% ...
\documentclass[a4paper,fleqn,usenatbib]{mnras}

\pdfoutput=1

\usepackage{newtxtext,newtxmath}

\usepackage[T1]{fontenc}
\usepackage{ae,aecompl}

\usepackage{graphicx}	
\usepackage{amsmath}	
\usepackage{amssymb}	

\newcommand{\Coftheta}{$C(\theta)$}
\newcommand{\Sonehalf}{$S_{1/2}$}
\newcommand{\lmin}{$\ell_{\rm min}$}
\newcommand{\pvalue}{$p$--value}
\newcommand{\pvalues}{$p$--values}
\newcommand{\LCDM}{$\Lambda$CDM}

\title[\Sonehalf\ non--anomaly]{The low level of correlation observed in the CMB sky at large angular scales and the low quadrupole variance}

\author[Knight and Knox]{
Robert Z Knight,$^{1}$\thanks{E--mail: riknight@ucdavis.edu (ZK)}
and Lloyd Knox,$^{1}$
\\
$^{1}$Physics Department, University of California, Davis, CA 95616
}

\date{Accepted XXX. Received YYY; in original form ZZZ}

\pubyear{2017}

\begin{document}
\label{firstpage}
\pagerange{\pageref{firstpage}--\pageref{lastpage}}
\maketitle

\begin{abstract}

The angular two-point correlation function of the temperature of the cosmic microwave background (CMB), as inferred from nearly all-sky maps, is very close to zero on large angular scales.  A statistic invented to quantify this feature, \Sonehalf, has a value sufficiently low that only about 7 in 1000 simulations generated assuming the standard cosmological model have lower values; i.e., it has a \pvalue\ of 0.007.  As such, it is one of several unusual features of the CMB sky on large scales, including the low value of the observed CMB quadrupole, whose importance is unclear: are they multiple and independent clues about physics beyond the cosmological standard model, or an expected consequence of our ability to find signals in Gaussian noise?  We find they are not independent: using only simulations with quadrupole values near the observed one, the \Sonehalf\ \pvalue\ increases from 0.007 to 0.08.  We also find strong evidence that corrections for a ``look-elsewhere effect'' are large.  To do so, we use a one-dimensional generalization of the \Sonehalf\ statistic, and select along the one dimension for the statistic that is most extreme. Subjecting our simulations to this process increases the \pvalue\ from 0.007 to 0.03; a result similar to that found in \cite{Pl16XVI}.  We argue that this optimization process along the one dimension provides an {\it underestimate} of the look-elsewhere effect correction for the historical human process of selecting the \Sonehalf\ statistic from a very high-dimensional space of alternative statistics {\it after} having examined the data.

\end{abstract}




\section{Introduction}

We view the continued success of a six-parameter model in describing the statistical properties of cosmic microwave background (CMB) temperature anisotropy maps with millions of resolution elements as one of the major takeaway messages from the {\it Planck} mission: a simple model works extremely well. This success includes agreement between model and data via multiple statistics, most prominently the two-point correlation function \citep{Pl14XV, Pl16XI}, but also via the higher-order correlations expected from gravitational lensing \citep{Pl14XVII, Pl16XV} and relativistic effects of our motion with respect to cosmic rest \citep{Pl14XXVII}. 

The strongest challenge\footnote{While others have described challenges to the success of the six-parameter model arising from some internal tensions revealed by tests using the two-point correlation function, e.g. \cite{Add16}, they are not highly significant \citep{Pl16LI}.}
to this takeaway message, based on CMB maps alone, may be the existence of relatively large-scale patterns that have been described as ``anomalous.''  Some of these large-scale CMB anomalies are: the existence of an unexpectedly large cold spot \citep{CTMV06,CCMVJ07,CMV10,ZH10}, the preference for odd parity modes \citep{KN10a,KN10b,KN11}, a hemispherical power asymmetry \citep{EHBGL04,EBGHL07,HEBGHL09,HBGEL09,AFSEHBG14,Pl14XXIII,Pl16XVI}, the alignment of the lowest multipoles with the geometry of the solar system and with each other \citep{dTZH04,HCMV04,SSHC04,LM05a,LM05b,Bea11,CHSS15b}, the low variance in low-resolution maps \citep{Pl14XXIII,Pl16XVI}, and the lack of correlation on the largest angular scales \citep{Bea03,Sp03,H07,CHSS07,CHSS09,Bea11,G14,CHSS15}.  For a recent review of large-scale CMB anomalies, see \cite{SCHS15}. 

One of the apparent anomalies that has received a lot of attention is the low value of the real space two-point correlation function at large angular scales. This feature can be seen in the earliest observations of the correlation function of the CMB as observed by the {\it Cosmic Background Explorer Differential Microwave Radiometer} ({\it COBE--DMR}) \citep{H96}, though it was not described in this way at the time.  The lack of correlations at large angles received more attention with the {\it Wilkinson Microwave Anisotropy Probe} ({\it WMAP}) first year data \citep{Sp03}, when the most widely used precise statistic created to capture this feature of the sky was defined: the \Sonehalf\ statistic.  \cite{Sp03} initiated the use of \Sonehalf, {\it after} having looked at the {\it COBE--DMR} and {\it WMAP} data, as a convenient way to show how unusual the lack of correlation of the CMB at large angles is under the assumption of \LCDM, making \Sonehalf\ an {\it a posteriori} statistic.  The \Sonehalf\ statistic has since been used to confirm the presence of the anomaly in the {\it WMAP} 3 year \citep{CHSS07} and 5 year \citep{CHSS09} data, as well as in the first \citep{CHSS15} and second \citep{Pl16XVI} {\it Planck} data releases.

We now define \Sonehalf, starting with a generalized version:

\begin{equation}
  	S_{x} = \int_{-1}^{x} \left[ C(\theta)\right]^2 d\cos(\theta)
\label{SonehalfEqn}
\end{equation}
  
\noindent where

 \begin{equation}
  	C(\theta) = \langle T(\mathbf{\hat n}) T(\mathbf{\hat n'}) \rangle
\label{autocorrEqn}
\end{equation}

\noindent is the two-point correlation function and $\theta$ is the angle between the two directions $\mathbf{\hat n}$ and $\mathbf{\hat n'}$.  With $x = 1/2 = \cos 60^\circ$, this becomes the standard definition of  \Sonehalf.  

Use of this statistic enables one to quantify how unusual the observed feature is, given \LCDM\ (the null hypothesis).  We use simulations to determine a distribution of \Sonehalf\ values given \LCDM, and find the fraction of simulations with lower \Sonehalf\ values than that inferred from the data.  This fraction of simulations that exceeds the statistic in question is commonly called a \pvalue.  
In a recent re-analysis of the observed lack of correlations on the CMB at large angles, \cite{CHSS15} calculated several \Sonehalf\ values of the CMB using various {\it WMAP} and {\it Planck} maps, masks, and methods, and using simulations based on the \LCDM\ theory, found \pvalues\ for \Sonehalf\ of the observed cut-sky CMB.  These vary from $0.00191$ to $0.00329$.  Alternatively, rather than use \pvalues, one could perform a Bayesian analysis on \Sonehalf\ in order to quantify how unexpected the observed value is.  \cite{EMH10} did such an analysis, in which they calculated an expected posterior distribution of theoretical \Sonehalf\ values, given the observed value.  This resulted in a very wide distribution of theoretical \Sonehalf\ values, and the \LCDM\ \Sonehalf\ value was not strongly disfavoured, indicating that \Sonehalf\ is actually a poor discriminator of theoretical models.  In this work, we do not use the Bayesian approach, but rather stick with the computation of \pvalues.

There have been numerous theoretical attempts to define cosmologies for which a low value of \Sonehalf\ is less unlikely.  Some of these attempts examine basic features of \LCDM, such as the Gaussianity and statistical homogeneity/isotropy of the primordial fluctuations. For example, \citet{CHSS09} showed that the low-$\ell$ modes of the CMB seem to work together to cause the lack of correlation on large angles, suggesting a correlation among the modes, indicating a violation of statistical isotropy.   Non-trivial topology could be a possible way to explain the low correlation on large angles, because in topological models that employ the spherical multi-connected manifolds, the power at large scales is naturally suppressed \citep{SSS93, NJ06}.  However, constraints on such manifolds arise from the null results of searches for matched circles on the CMB.  In a more recent work, \cite{AL14} explore CMB correlations in a cosmology incorporating the Hantzsche--Wendt manifold, which is interesting because this manifold could easily escape detection by matched circle pairs.  They calculate the distribution of \Sonehalf\ for an ensemble of simulations in this topology, finding it reduced by about a factor of 2 compared to \LCDM.  

Here we argue that this attention is misplaced. We find the evidence is quite weak that the low \Sonehalf\ value is due to anything other than either the low quadrupole value, or a human's ability to identify unusual-looking features in a high-dimensional data set with many different ways it can be unusual, or both.  

The role of the low-order multipoles and the quadrupole in particular in the lack of correlation at large angles has previously been investigated.  \cite{BVWLF06} examined the role of the low quadrupole in the correlation function by removing the quadrupole from both the observed {\it WMAP} data and the \LCDM\ model, and found a qualitatively better agreement between observation and theory than with the quadrupole included.  Furthermore, both the quadrupole-removed {\it WMAP} and \LCDM\ correlation functions showed a lack of correlations at large angles, whereas the non-quadrupole-removed \LCDM\ correlation function did not, indicating that the low quadrupole is largely, though not entirely, responsible for the lack of correlation at large angles. \cite{H07} pointed out that the contribution to \Coftheta\ from multipoles $\ell = 2,3$ in the observed CMB is nearly equal and opposite to the contribution from the rest of the multipoles put together. Additionally, \cite{CHSS09} found that by tuning the powers $C_2, C_3$ of the \LCDM\ prediction, they could obtain an \Sonehalf\ expectation value lower than the observed \Sonehalf\ value found from {\it WMAP}.  However, no work of which we are aware has yet reconciled the low \pvalue\ of the \Sonehalf\ statistic at $\sim0.007$, with the larger \pvalue\ of the low quadrupole at $\sim0.04$.  

Despite this previous attention to the relationship between the low quadrupole and low correlation at large angles, we find an important result has been overlooked: namely, the impact of the low quadrupole on the distribution of \Sonehalf\ values expected under \LCDM. We demonstrate that once the probability of \Sonehalf\ is conditioned on the observed value of the quadrupole, the value of \Sonehalf\ is no longer anomalous, as the \pvalue\ rises to 0.08. This result has implications for solutions to the puzzle of low \Sonehalf: any change to the model that suppresses the ensemble average of the quadrupole variance is helpful.  For example, several groups have explored the possibility of an era in the early universe before slow-roll inflation in which a ``fast-roll'' caused a large scale suppression of power, effectively creating a cutoff in the primordial power spectrum which would be observable as a lack of correlations on large scales today \citep{CPKL03, LBH14, LGP14, GS14}.  

Although the low value of \Sonehalf\ is potentially understandable as due to a departure from \LCDM\ that reduces the ensemble average of the quadrupole, the question still remains of whether \Sonehalf\ is compatible with \LCDM\ in the absence of any such modifications.

The standard criticism of  the amount of attention given to the anomalies is the standard criticism of {\it a posteriori} statistics: their {{\it a posteriori} nature makes their interpretation very challenging. Consider that for a map of the sky, even with a finite number of pixels or data points, there is an infinite variety of statistics that one can make up.\footnote{As a simple example, consider linear combinations of a finite number of data points.  Given the freedom to choose arbitrary coefficients, there are an infinite number of possible combinations of these data.} If one makes up 1,000 of them prior to examining some data set, then it is not evidence of a failure of a model if one of those statistics has a value that happens to be that extreme in only 1 out of 1,000 simulations. With {\it a posteriori} statistics, by definition, one creates the statistic {\em after} seeing the data. When such a statistic ends up having an extreme value, determining whether or not this indicates a failure of a model is difficult because it is not clear how many such statistics could have been made up, had the data looked different to begin with. Additionally, because the choice of statistic to use was informed by the data, this also makes simulation of the data creation and analysis process very difficult, and perhaps practically impossible. What statistic would a different, simulated data set, have led the analyst to choose? Now that an analyst is in the simulation loop, simulation is quite challenging, to say the least. 

There are several features about the \Sonehalf\ statistic which can be considered to have been selected {\it a posteriori}.  Broadly speaking, the whole notion that \Coftheta\ is low at large angles is {\it a posteriori}.  More specifically, the choices to square \Coftheta, to integrate this quantity over angles, and to select an upper integration limit of $1/2$ were all {\it a posteriori} choices, thereby making the significance of the value of \Sonehalf\ difficult to interpret.  In order to address the critique that the upper integration limit was selected {\it a posteriori}, the {\it Planck} team \citep{Pl16XVI} recently implemented a process in which this limit in the definition of \Sonehalf\ was allowed to vary over the whole range of $x$ values $(-1 \le x \le 1)$. We repeat a similar process here, although with a different rationale. We know historically that there actually was no optimization over $x$ as a continuous variable -- this point is indisputable as one can do the optimization and the result is not $x=1/2$. However, the choice of a statistic tailored to capture the apparently unusual near-zero values of the correlation function on large angular scales was clearly an {\it a posteriori} selection. We view the $S_x$ statistic as a one-dimensional proxy for the very high-dimensional space of all possible statistics from which $S_{1/2}$ was selected. It is a useful proxy as the impact of {\it a posteriori} choice can then be calculated. When we consider the effects of the {\it a posteriori} choice of the \Sonehalf\ statistic, by considering other possible statistics that one could calculate, the significance of the \Sonehalf\ anomaly decreases.  However, as we are using a one-dimensional proxy, we expect the impact of selection from the larger space is underestimated. \footnote{For completeness, we mention that \cite{EMH10}, in addition to providing the Bayesian analysis we mentioned earlier, also suggested that a two-dimensional generalization of \Sonehalf\ could illustrate the impact of {\it a posteriori} choice on \pvalues\, but did not make any clear quantitative claims about this impact other than to suggest it could be ``an order of magnitude or more.''}

One research direction motivated by the \Sonehalf\ anomaly that we do see as valuable is a search for testable predictions of the hypothesis that \LCDM\ is correct and the low value of \Sonehalf\ is just a statistical fluke. \cite{YCSK14} use \LCDM\ realizations conditioned on the low value of \Sonehalf\ to develop {\it a priori} statistics for use with future data sets, such as CMB polarization data, designed to test the hypothesis that the low \Sonehalf\ measurement is just the consequence of an unlikely \LCDM\ realization. The conclusion of our work here is that the evidence is quite weak that the low \Sonehalf\ value indicates anything other than a somewhat unlikely \LCDM\ realization. We hesitate to call it a fluke however, as it is only a highly unusual realization if one does not correct for {\it a posteriori} effects. Names aside, we expect that fluke-hypothesis tests have the potential to confirm the ``fluke'' interpretation we favor here, or provide stronger evidence against \LCDM. 

The rest of this paper is organized as follows.  In section \ref{SonehalfStat} we describe the observational and simulated data and data processing needed to create \Sonehalf\ \pvalues.  In section \ref{lowQuadrupole} we describe the effect of the low quadrupole on the observed \Sonehalf\ value and how we include the low quadrupole in our filtered-\LCDM\ null hypothesis, and the effect that that has on the \Sonehalf\ \pvalue.  In section \ref{noAPosteriori} we describe a means to quantify the impact of the {\it a posteriori} choice $x = 1/2$ by exploring a more generic statistic $S_x$.  We finish by combining these two methods and examining the resultant quadrupole-filtered look--elsewhere--corrected \pvalues.

\section{The \Sonehalf\ Statistic}\label{SonehalfStat}

In order to calculate how unlikely the \Sonehalf\ statistic for the observed CMB is, given a particular cosmological model, one has to calculate \Sonehalf\ for the observed universe as well as for an ensemble of simulated universes, created based on that model.  When compared to a suitably created ensemble of simulations, the \pvalue\ of the \Sonehalf\ statistic can be computed.  In this analysis, we have treated the observed CMB map as similarly as possible to those created in simulations, for consistency.  This includes the treatment of the pixelization, spatial frequency content, and masking.

\subsection{CMB Observations}\label{CMBobs}

There are to date several (nearly) full-sky maps of the CMB that are suitable for use in calculating the correlation function $C(\theta)$, and the corresponding large angle statistic \Sonehalf.  \citet{CHSS15} used two different Galactic+foreground masks\footnote{A mask is a set of flags indicating which pixels of a map are to be excluded from use in further analysis.} in their study: the {\it WMAP} 9-year \texttt{KQ75y9} mask, which leaves unmasked $69$ per cent of the sky ($f_{\rm sky} = 0.69$), and a {\it Planck}-derived \texttt{U74} mask, which has $f_{\rm sky} = 0.74$.  They used each of these masks with several {\it WMAP} and {\it Planck} CMB maps, and in every case, the \Sonehalf\ \pvalues\ were considerably higher (meaning the observed \Sonehalf\ was less unusual) for the smaller \texttt{U74} mask (larger $f_{\rm sky}$) than those for the larger \texttt{KQ75y9} mask (smaller $f_{\rm sky}$).  This is also consistent with the results of \cite{G14}, who showed that larger masks (smaller $f_{\rm sky}$) make the significance of the \Sonehalf\ anomaly larger (smaller \pvalue) than with smaller masks.  With an array of Galactic masks that ranged in sizes from $f_{\rm sky}=0.78$ to $0.28$, they found a corresponding decrease in \pvalue\ from $1.96$ per cent to $0.05$ per cent.  (Note that within this range, the two mask used by \cite{CHSS15} were on the smaller end, or had larger $f_{\rm sky}$.)  Except where otherwise mentioned, in this analysis we use only one observed CMB map and mask combination: the {\it Planck} PR2 (2015) \texttt{SMICA} map with the (2015) \texttt{Common} mask.\footnote{\label{pl_archive} available at the {\it Planck} Legacy Archive \url{http://pla.esac.esa.int/pla/\#home}}  

The {\it Planck} map and mask are stored in the \textsc{healpix}\footnote{\label{healpixFootnote} See \url{http://healpix.sourceforge.net} for information about \textsc{healpix}} \citep{healpix05} format at high angular resolution.  For calculations of \Sonehalf, high angular resolution is not needed, so we degrade the resolution of the mask and map from $N_{\rm side} = 2048$ to $N_{\rm side} = 128$ following the procedure described by \cite{Pl16XVI}:  the map is first transformed to harmonic space using \textsc{healpix}, then a reweighing by the ratio of pixel window functions is applied, and then the harmonic coefficients are re-transformed back into a map at the lower resolution.  The same thing is done to the mask, with the one additional step of defining a threshold value of 0.9, over which the mask pixel values are set to 1, and below which they are set to 0.  The resulting mask leaves a fraction of the sky $f_{\rm sky} = 0.676$ available for analysis.  The low resolution of $N_{\rm side} = 128$ was chosen in order to reduce computational time while preserving the signal of interest, which is at large angular scales, and also to be consistent with the resolution used in previous analyses (e.g. \cite{CHSS15}).

The power spectrum $C_\ell$ of the degraded \texttt{SMICA} map was calculated using \textsc{spice}.\footnote{\label{spiceNote} \textsc{polspice} available at \url{http://www2.iap.fr/users/hivon/software/PolSpice/}}  This code applies the mask to the map and subtracts the monopole and dipole from the resultant cut-sky map as part of the process to find the cut-sky $C_\ell$.  To derive the correlation function \Coftheta\ shown in Fig.~\ref{fig:autocorrLmin2}, we applied the transform

\begin{equation}
	C(\theta) \equiv \sum_{\ell=2}^{100} \frac{2\ell +1}{4\pi} C_\ell P_\ell(\cos \theta).
\label{legTransIsotropic}
\end{equation}

\noindent  We used the upper limit of 100 following \cite{CHSS15}, who showed that the effect on \Sonehalf\ of using $\ell_{\rm max} = 100$ rather than a higher $\ell$ value was less than 1\%.

In order to calculate \Sonehalf\ for the \texttt{SMICA} map, we used the harmonic space method of \cite{CHSS09}.  Combining equations \ref{SonehalfEqn} and \ref{legTransIsotropic}, we evaluate  

\begin{equation}
	S_{x} = \sum_{\ell,\ell'} C_\ell I_{\ell,\ell'}(x) C_{\ell'}
\label{matrixSonehalf}
\end{equation}

\noindent where the quantity 

\begin{equation}
I_{\ell,\ell'}(x) = \frac{(2\ell+1)(2\ell'+1)}{(4\pi)^2} \int_{-1}^{x} P_{\ell}(\mu)P_{\ell'}(\mu) d\mu
\label{IofxEqn}
\end{equation}

\noindent is an easily calculable function of $x$ \citep{CHSS09}, and we have used the generalization to $S_x$, with \Sonehalf\ being a special case with $x = 1/2$.

Using our degraded \texttt{SMICA} map, we calculated \Coftheta\ and \Sonehalf\ resulting in a value of  $S_{1/2} = 2145~{\rm \mu K}^4$, which when compared to our ensemble of $10^5$ simulations, has a \pvalue\ of $0.72$ per cent  (see the solid curves in Fig.~\ref{fig:autocorrLmin2} and Fig.~\ref{fig:SonehalfLmin}).  This \Sonehalf\ value is quite close to the value $S_{1/2} = 2153~{\rm \mu K}^4$ found by \cite{COS16}, who used the same data set and methods, and is similar to those found by \cite{CHSS15}, who used older {\it Planck} and {\it WMAP} data sets.  The \Sonehalf\ \pvalues\ reported by \cite{CHSS15} range from about $0.2$ per cent to $0.3$ per cent, depending on the map, mask, and method used.  For a more detailed comparison, see the appendix.

\subsection{Creating Simulations}\label{makeSims}

Simulations of the CMB were created using \textsc{healpix} based on a theoretical power spectrum created by \textsc{class}\footnote{\label{classNote} \textsc{class} available at \url{http://class-code.net/}}.  To create this power spectrum, we used the best fit 2015 {\it Planck} \LCDM\ parameters \citep{Pl16XIII}:  $h=0.6774$, $\Omega_{\rm b} = 0.04860$, $N_{\rm eff} = 3.04$, $\Omega_{\rm cdm} = 0.2589$, $\Omega_\Lambda = 0.6911$, $Y_{\rm p}=0.249$, and $z_{\rm reio}=8.8$, along with the {\it WMAP} value \citep{F09} $T_{\rm cmb} = 2.726~{\rm K}$.  Prior to creating realizations, the power spectrum was dampened slightly with pixel window and Gaussian beam functions, since the observational data, as presented, contains pixel window and telescope beam effects.

The \textsc{healpix} simulations were then created at the same resolution ($N_{\rm side} = 128$) as our degraded-resolution \texttt{SMICA} map, with the same harmonic content ($2 \leq \ell \leq 100$) that is required to perform the \Sonehalf\ calculation in equation \ref{matrixSonehalf}.

Each simulation was treated in the same way as the \texttt{SMICA} map: we fed each simulation into \textsc{spice}, which applied the degraded-resolution \texttt{Common} mask and removed the monopole and dipole in order to deliver an estimate of the power spectrum $C_\ell$.  We created two separate ensembles of power spectra: one with $10^4$ and one with $10^5$ members.  The spectra in the smaller set were transformed via equation \ref{legTransIsotropic} in order to create the ensemble average autocorrelation function \Coftheta\ and $1\sigma$ band, shown in Fig.~\ref{fig:autocorrLmin2}.  Those in the larger set were used directly in equation \ref{matrixSonehalf} with $x = 1/2$ in order to create an ensemble of \Sonehalf\ values, shown in Fig.~\ref{fig:SonehalfLmin}, with 3 different values of \lmin, the lowest multipole included in the summation.

\begin{figure}
\includegraphics[width=\columnwidth]{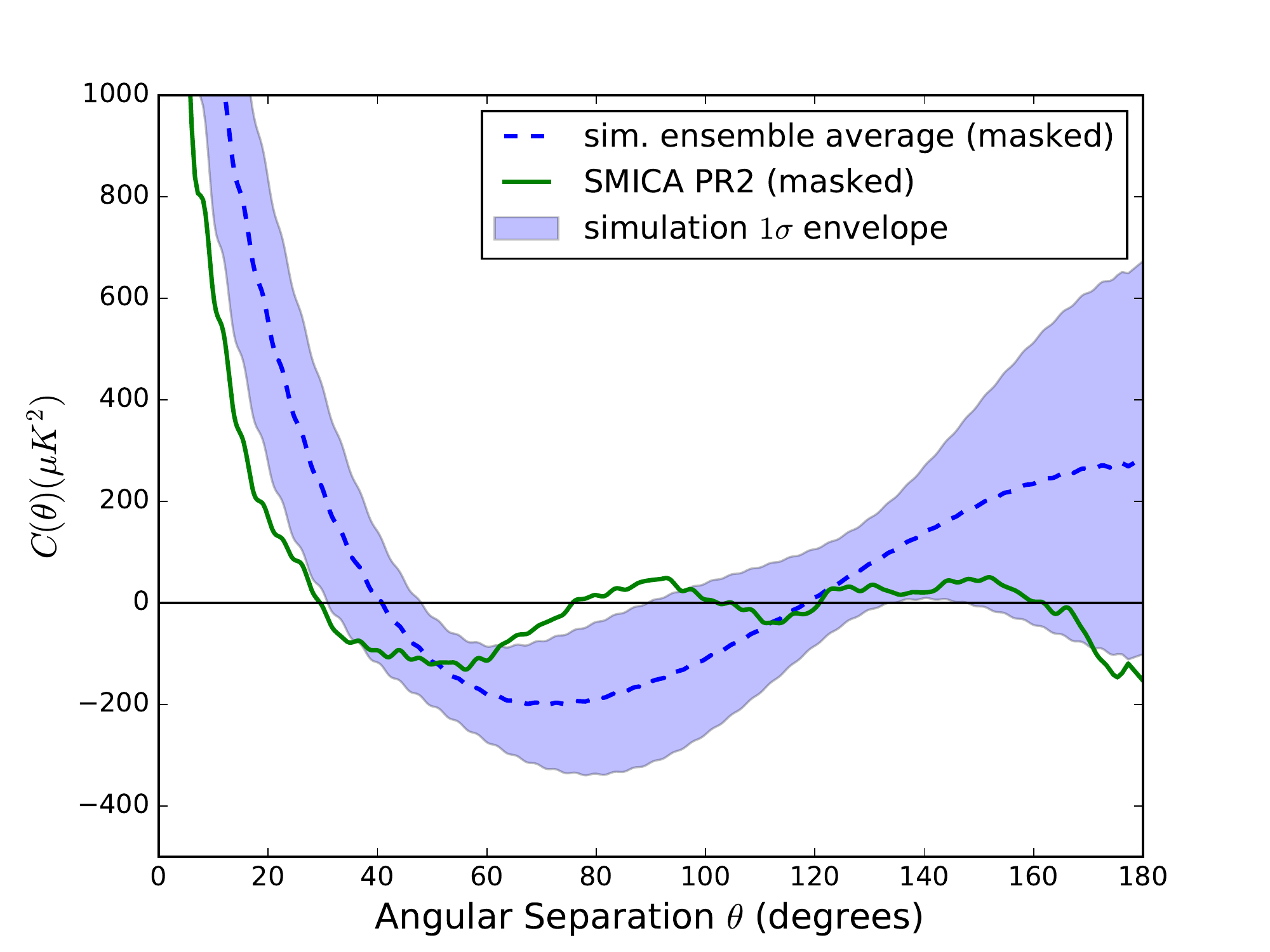}
\caption{The auto-correlation $C(\theta) = \langle T(\mathbf{\hat n}) T(\mathbf{\hat n'}) \rangle$ where $\cos\theta = \bold{\hat n} \cdot \bold{\hat n'}$, of the masked \texttt{SMICA} map (solid green line), with the average and $1\sigma$ confidence region of the autocorrelation functions of $10^4$ masked CMB simulations (dashed and shaded blue).}
\label{fig:autocorrLmin2}
\end{figure}


\section{Effect of the Low Quadrupole Power}\label{lowQuadrupole}

The summation formulae for \Coftheta\ and \Sonehalf\ (equations \ref{legTransIsotropic}, \ref{matrixSonehalf}) make it easy to examine the relative importance of each multipole to \Coftheta\ and \Sonehalf.  We found that removing the lowest multipoles from the calculation of \Sonehalf\ for both simulations and for the \texttt{SMICA} data drastically increased the \pvalue\ of the \texttt{SMICA} \Sonehalf\ statistic (See Fig.~\ref{fig:SonehalfLmin}).  In fact, for the \lmin\ $= 4$ case, the \texttt{SMICA} value is quite near the middle of the simulated distribution.  This shows that the low \pvalue\ of the standard \Sonehalf\ statistic is highly influenced by the $C_2$ and $C_3$ of the observed CMB.

\begin{figure}
\includegraphics[width=\columnwidth]{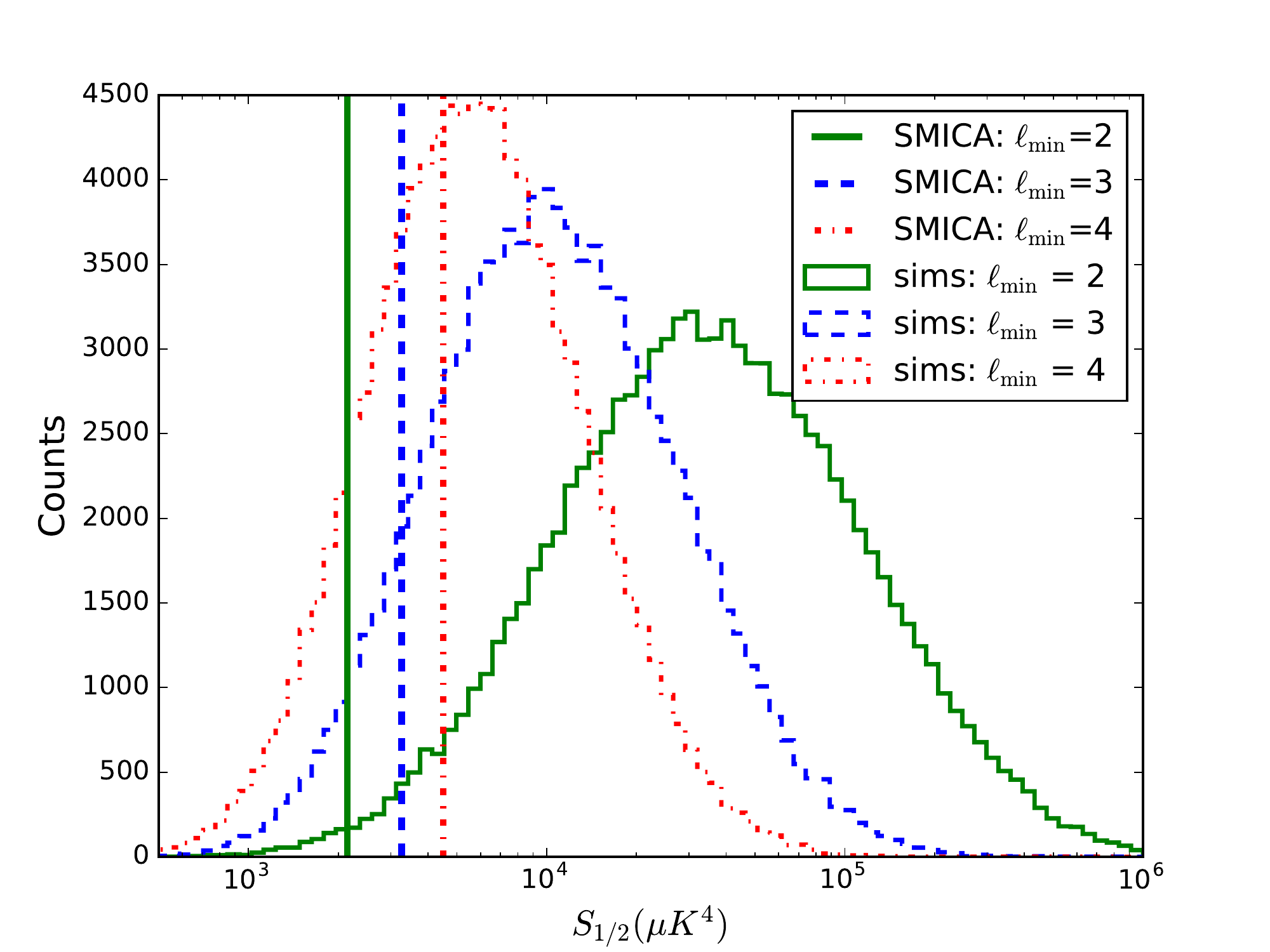} 
\caption{\Sonehalf\ calculated for the masked \texttt{SMICA} map (vertical lines) and for $10^5$ masked simulations (histograms).  The minimum $\ell$ value included in the calculation is varied.  Note that for these values, as \lmin\ increases, the \texttt{SMICA} \Sonehalf\ value increases, while the ensemble values tend to decrease. The \pvalues\ for the ensembles shown here are $0.72$ per cent, $10.6$ per cent, and $38.7$ per cent. }
\label{fig:SonehalfLmin}
\end{figure}

Here we revisit the question of the connection between $\ell = 2$ and \Sonehalf.  In particular, we ask how dependent the \Sonehalf\ anomaly is on the low observed value of $C_2$.  In our \LCDM\ model, the ensemble average quadrupole variance is $1099.94~{\rm \mu K}^2$, whereas for the degraded and masked \texttt{SMICA} map, we measure a quadrupole variance of only $171.8~{\rm \mu K}^2$, which is lower than the ensemble average value by a factor of 0.156.  However, the low observed quadrupole variance is by itself not exceedingly anomalous.  To see this, we created an ensemble of $10^5$ simulated CMB skies, masked them, and measured their $C_2$ and \Sonehalf\ values (see Fig.~\ref{fig:C2vsSonehalf}).  In this ensemble, the cut-sky \texttt{SMICA} $C_2$ has a \pvalue\ of $3.9$ per cent, which is much higher than the \Sonehalf\ \pvalue\ of $0.72$ per cent.  This is consistent with a similar analysis by \cite{E03}, who calculated $C_2$ \pvalues\ using two differently measured {\it WMAP} $C_2$ values of $129$ and $212~{\rm \mu K^2}$,\footnote{The variances presented by \cite{E03} were given in terms of $\Delta T^2_\ell = \ell(\ell+1)C_\ell/(2\pi)$.} which resulted in \pvalues\ of $1.3$ and $3.6$~per~cent, respectively.  This suggests, as \cite{CHSS09} point out, that the low quadrupole variance anomaly is not the same as the \Sonehalf\ anomaly.  

This joint \Sonehalf$-C_2$ distribution allows us to look at the ensemble relationships between the two values.  As can be seen in the scatter plot and density contours of Fig.~\ref{fig:C2vsSonehalf}, the distribution has a golf-club like appearance, such that the upper right of the plot shows a thin and narrow structure, while the lower left is wedge shaped.  The thin distribution in the upper right indicates that at high values, \Sonehalf\ and $C_2$ are correlated with each other.  However, this correlation breaks down at lower values, in the wedge-shaped part of the distribution.  At these low values, a low value of \Sonehalf\ indicates that $C_2$ must also be low, but the inverse, that a low value of $C_2$ indicates a low value of \Sonehalf, is not true.  However, as we show below, conditioning an ensemble to only include low values of $C_2$ does have a significant impact on the distribution of \Sonehalf\ values, and increases the \texttt{SMICA} \pvalue\ substantially.

\begin{figure}
\includegraphics[width=\columnwidth]{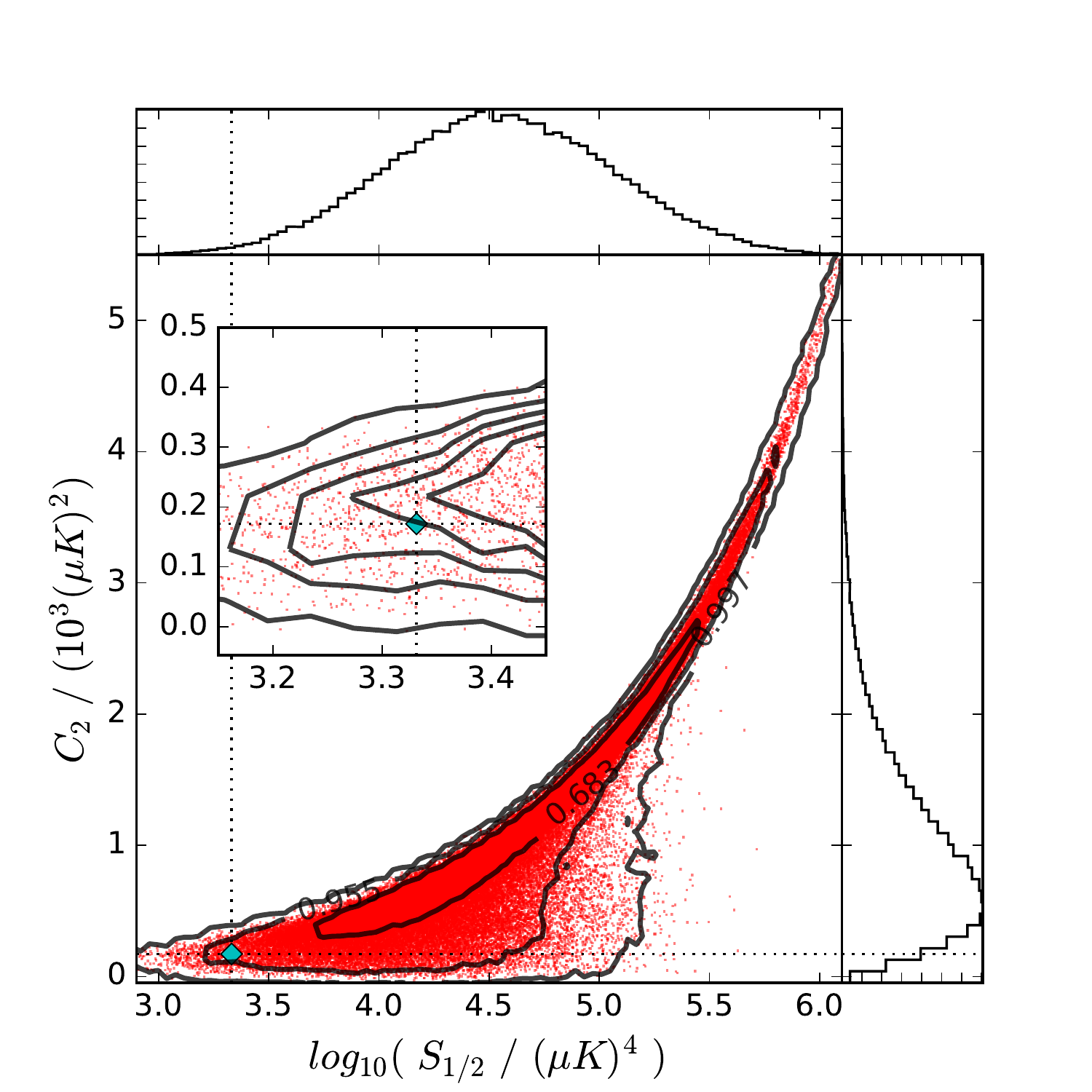} 
\caption{$C_2$ and \Sonehalf\ values for $10^5$ masked-sky CMB simulations, with density contours containing $68.3$, $95.5$, and $99.7$ per cent of the points.  Marginalizations are shown in the marginal plots.  Inset is a zoom of the area around the \texttt{SMICA} point, with $91$, $93$, $95$, $97$, and $99$ per cent density contours shown.  The \texttt{SMICA} value is marked by the blue diamond in each plot; it is within the $95.5$ per cent contour on the main plot, and just slightly within the $93$ per cent contour on the inset.  The location of the diamond near the lower corner of the distribution indicates the low \pvalues\ of \Sonehalf\ and $C_2$.   The \Sonehalf\ distribution (with \lmin\ $=2$) shown in Fig.~\ref{fig:SonehalfLmin} can be found by marginalizing over $C_2$, whereas the constrained \Sonehalf\ distribution shown in Fig.~\ref{fig:SonehalfFiltered} can be thought of as a narrow horizontal slice through the location of the \texttt{SMICA} mark.}
\label{fig:C2vsSonehalf}
\end{figure}

We implemented a constraint on the quadrupole variance of our cut-sky simulations by generate--and--test filtering.  We first created a simulation, masked it, then measured its quadrupole variance and compared it to the \LCDM\ expectation value.  If $C_2$ of the simulation fell between 0.1 and 0.2 times the \LCDM\ expectation value (where 0.1 and 0.2 were chosen to simply bracket the observed ratio of 0.156 with round numbers), it was included in the ensemble of simulations.  If not, we threw it out.  We kept creating simulations until we had reached the desired number in the ensemble: $10^4$.

The results of this ensemble selection process are shown in Fig.~\ref{fig:autocorrFiltered} and Fig.~\ref{fig:SonehalfFiltered}.  Comparing these to the previous versions without filtering (Fig.~\ref{fig:autocorrLmin2} and Fig.~\ref{fig:SonehalfLmin}), we see that the ensemble of correlation functions has changed shapes considerably, converging toward the line indicating zero, and the \Sonehalf\ distribution has correspondingly lowered toward zero.  The \texttt{SMICA} \Sonehalf\ value (unchanged) therefore appears to be in a much less unlikely place in this distribution, with a \pvalue\ of $8.24$ per cent.  This non-anomalous \pvalue\ suggests that, at the very least, \Sonehalf\ is not independent from the low quadrupole.  Once we condition on the observed quadrupole value, the observed \Sonehalf\ value is not that rare.

\begin{figure}
\includegraphics[width=\columnwidth]{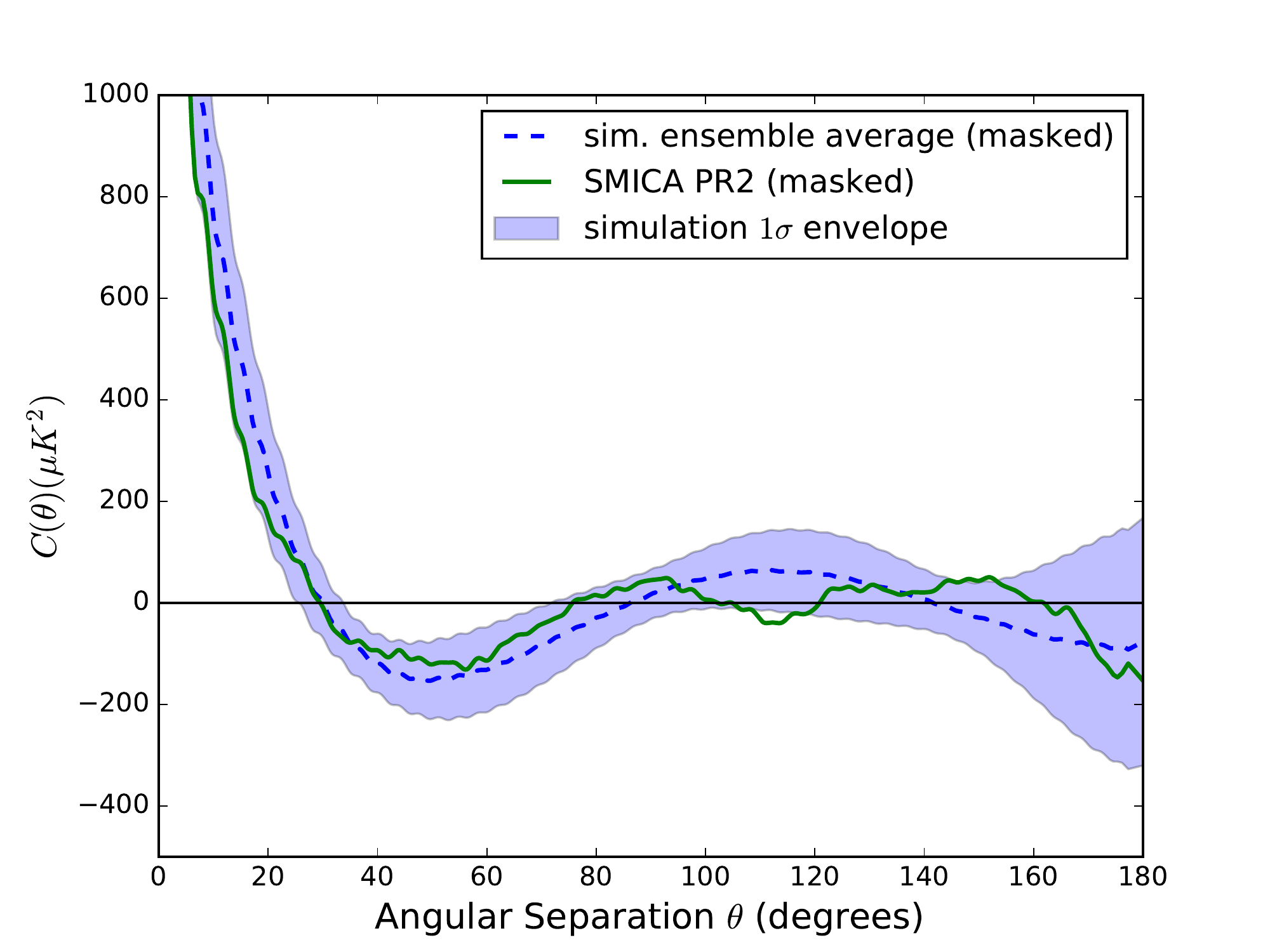}
\caption{The auto-correlation as in Fig.~\ref{fig:autocorrLmin2}, but with simulations selected for the ensemble only if  $0.1\times C_2^{\rm \Lambda CDM} < C_2^{\rm sim} < 0.2\times C_2^{\rm \Lambda CDM}$. }
\label{fig:autocorrFiltered}
\end{figure}

\begin{figure}
\includegraphics[width=\columnwidth]{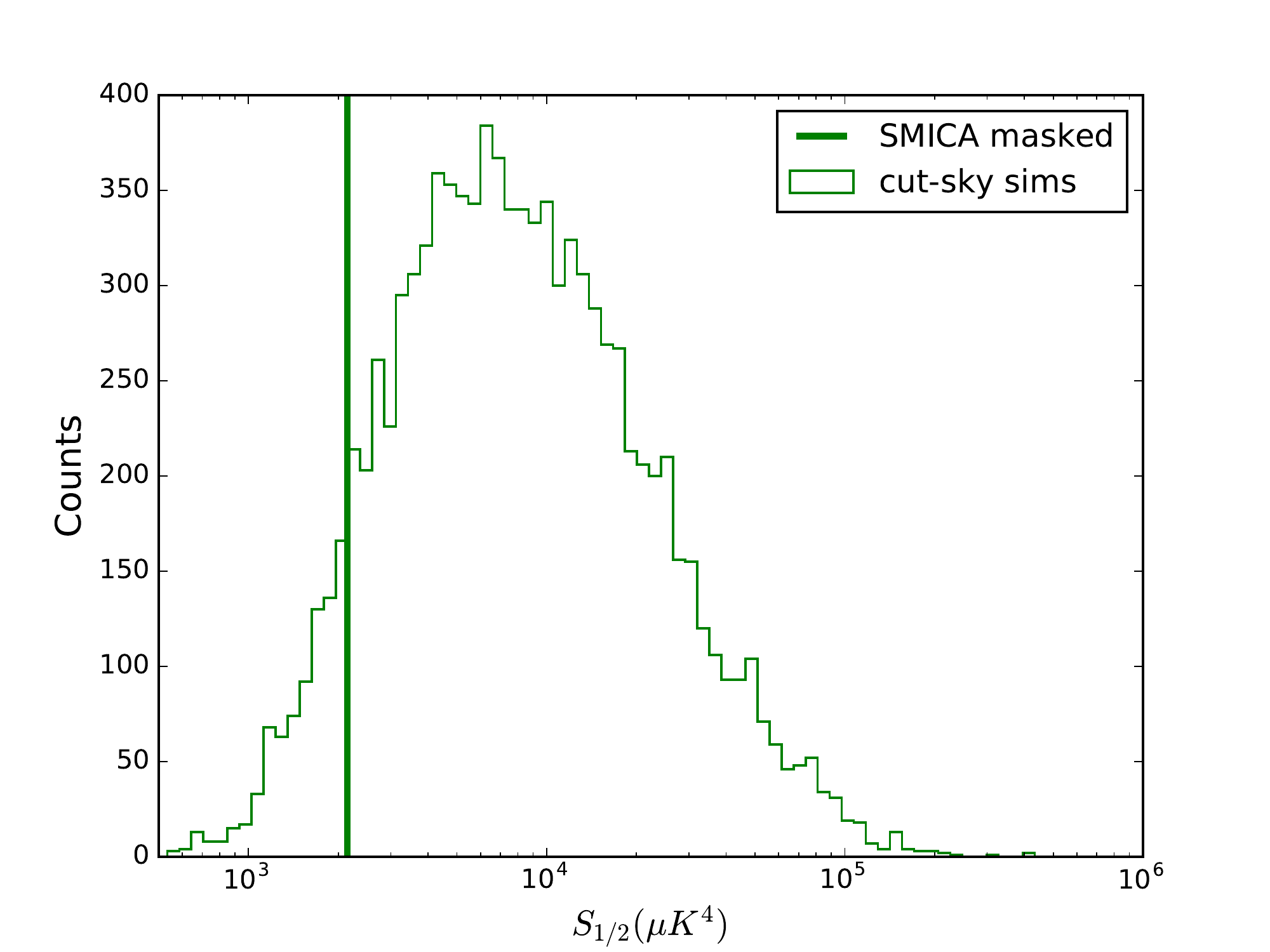}
\caption{\Sonehalf\ as in Fig.~\ref{fig:SonehalfLmin} (\lmin\ $=2$), but with simulations selected for the ensemble only if  $0.1\times C_2^{\rm \Lambda CDM} < C_2^{\rm sim} < 0.2\times C_2^{\rm \Lambda CDM}$, as in Fig.~\ref{fig:autocorrFiltered}.  In this $C_2$-filtered ensemble of $10^4$ simulations, \Sonehalf\ has a \pvalue\ of $8.24$ per cent Note that this is slightly lower than the quadrupole-removed \pvalue\ of $10.6$ per cent, indicated by the \lmin\ $=3$ case in Fig.~\ref{fig:SonehalfLmin} } 
\label{fig:SonehalfFiltered}
\end{figure}

Finally, we look back at the joint \Sonehalf$-C_2$ distribution, and generalize the notion of the \pvalue\ to this 2 dimensional space.  To do so, we created iso-probability density contours by approximating the probability density using kernel density estimation.  Some of these contours are shown in Fig.~\ref{fig:C2vsSonehalf}, labelled by what fraction of the total number of points they contain.  The \texttt{SMICA} value lies in a region which has a probability density that is not extremely low, lying just within the $93$ per cent contour, indicating a 2D \pvalue\ of $7$ per cent, less than a $2\sigma$ deviation from the densest region.  To compare this to the 1D cases we need to consider that the definition of \pvalue\ that we have been using is only a 1-tailed statistic, and does not account for outliers that are extreme in the opposite direction, at equal or lower probability density.  (Our density contours in 2D account for high and low values in all directions.)  Therefore, to approximate an equivalent 2-tailed statistic, we multiply the 1-tailed 1D \pvalues\ by 2, giving us 2-tailed 1D \pvalues\ of $1.44$ per cent for \Sonehalf, and $7.8$ per cent for $C_2$, which is clearly an underestimate given the asymmetry of the 1D $C_2$ distribution.  Thus we see that the observed values in the joint space occupy a probability density that is similar, in terms of how extreme it is, to the one-dimensional case of $C_2$. We conclude that there is no new evidence in this joint space for anomalous behavior.

\section{Effect of the a Posteriori Choice}\label{noAPosteriori}

The previous section provides a better description of the influence of the low quadrupole on the lack of correlations at large angles then was available before, suggesting that modifications to \LCDM\ which lower the expected quadrupole contribution to \Coftheta\ would help reduce the anomalous nature of \Sonehalf.  However, we are still interested in whether \Sonehalf\ is consistent with \LCDM\ without any such modification.  To that end, we now turn to addressing what is perhaps the most often criticized aspect of \Sonehalf: that it was created to capture a feature of the data by analysts who had already taken a look at the data.  That is, \Sonehalf\ was created {\it a posteriori}.\footnote{For a {\it reductio ad absurdum} description of {\it a posteriori} statistic selection in the context of CMB and $\pi$ anomalies, see \cite{FS16}.}  

If, prior to looking at the data, one came up with 1,000 statistics to calculate from the data, and then applied them to the data, assuming they were all uncorrelated, one would expect a uniform distribution of \pvalues\ between 0 and 1. The expectation value for the number of statistics with \pvalues\ between 0 and 0.001 would be 1. Even if they were correlated, one could still perform simulations that would naturally take these correlations into account, and use them to calculate the probability that one of the statistics returned a \pvalue\ of less than 0.001, by counting the fraction of simulations for which this is the case; that is, we calculate the \pvalue\ of the \pvalue. Such a probability we call the look--elsewhere--corrected (LEC) \pvalue.

To perform the analogous calculation for a single {\it a posteriori} statistic, one chosen after inspection of the data, the simulation process would include the process of inspecting the data and inventing a statistic to capture what is perceived to be an unusual feature. This is obviously challenging, if not impossible, as it requires an automated model of the analyst him or herself, so that the simulated data are processed by the simulated analyst who identifies the unusual feature and creates a statistic designed to capture that feature in a single number.  

The difficult aspect of the above procedure is the automation of the human process of selection of a statistic from a very high-dimensional array of possibilities. In the following we adopt a crude, but calculable, approximation of this process as the selection of one statistic from a merely one-dimensonal space of alternatives. We create this one dimensional space by extending one single aspect of the already well defined \Sonehalf\ statistic: the choice of the upper endpoint of the \Sonehalf\ integral, $1/2$.  As described above, we have generalized \Sonehalf\ to $S_x$, which has an arbitrary upper integration limit, chosen from the whole range of possible values: $-1 \le x \le 1$.

Note that we are not claiming that this is the procedure that was followed historically. We know it is not because, as we will see, the choice of $x$ that returns  the most extreme vaue of $S_x$ from the real data is $x=0.37$ rather than $x=1/2$. The resulting LEC \pvalue\ we obtain should be viewed as merely indicative of the type of correction that is plausible. We suspect that, if anything, our replacement of the high-dimensional space of alternative statistics with a one-dimensional one {\em under}estimates the size of the correction.

We proceed by using the generalized $S_x$ integral (equation~\ref{SonehalfEqn}) and calculating $S_x$ as a function of $x$ over the whole range of $x$ values for an ensemble of $10^5$ CMB simulations (without $C_2$ filtering this time).  Then, to save computational time, we choose a subsample of these, the first $10^4$ curves, to evaluate further.  For each $S(x)$ function in the subsample we find, for each value of $x$, a rank in the entire set of $10^5\ S_x$ curves.  We divide these ranks by $10^5$ and they become (1-tailed) \pvalues, as a function of $x$: $p(x)$.  Then, for each simulation in the subsample, we find its minimum \pvalue\ and corresponding $x$ and $S_x$ values.  In order to treat the \texttt{SMICA} data in the same manner, we throw it in as just another member of this ensemble and find its optimal $x$, $S_x$, and $p(x)$ values with resect to the ensemble as well.  Finally, using the ensemble of \pvalues, we can calculate the LEC \pvalue.

The result of this process is shown\footnote{\label{cornerNote} plotting package corner.py available at \url{https://github.com/dfm/corner.py}} in Fig.~\ref{fig:cornerplotNofilter}, where the \texttt{SMICA} value is indicated by the blue horizontal and vertical lines.  In this ensemble, the \texttt{SMICA} map, optimized for minimum \pvalue, has $x = 0.37$,  $S_x = 1326~{\rm \mu K}^2$, and a nominal $p(x) = 0.36$ per cent.  This is lower than the \Sonehalf\ \pvalue\ of $0.72$ per cent that we calculated earlier, which is as expected due to the optimization process.  The LEC \pvalue\ is $2.94$ per cent.

To calculate the error on the LEC \pvalue, we estimated the sample variance of these quantities by doing some calculations with smaller $S_x$ ensembles.  These smaller ensembles were constructed as subsets of the same set of $10^5\ S_x$ simulations that we used before, but with $10^4$ members chosen randomly, and with $p(x)$ curves calculated using only the corresponding $10^4\ S_x$ curves, rather than the entire set of $10^5$ of them, as we did previously.\footnote{This is a variation on the jackknife resampling test.}  Interestingly, these fell into two groups, with 27 of them having $x$ values near 0.35, and 5 of them being near -0.1.  For those near $x = 0.35$, the average \pvalue\ was $0.330$ per cent, and for those near $x = -0.1$, the average \pvalue\ was $0.346$ per cent.  There does not appear to be a big difference in \pvalues\ between the two groups, but since our first sample had an $x$ value of 0.37, we compare it to the first group.  Supposing that the sample variance is dominated by the number of $S_x$ curves used in the calculation, we expect the sample variance of these \pvalues\ to be approximately equal to the Poisson noise associated with the number of simulations used.  Since the average \pvalue\ of $0.33$ per cent corresponds to 33 out of 10000 samples, we hypothesized that these 27 optimal \pvalues\ were drawn from a Poisson distribution with mean 33, and performed a Kolmogorov--Smirnov goodness-of-fit test.  This resulted in a Kolmogorov--Smirnov \pvalue\ of 0.24, suggesting that these values are consistent with being drawn from a Poissonian distribution, as hypothesized.  Therefore, we also suppose that the \Sonehalf\ \pvalue\ that we obtained using $10^5\ S_x$ curves, also subject to a sample variance, can also be thought of as being drawn from a Poissonian distribution.  Using the only \pvalue\ that we have ($0.361\% = 361/10^5$) to derive an estimate of the mean, we find the square root of the Poissonian variance to be $\sqrt{361} = 19$, constraining our measured \pvalue\ to be $0.361\% \pm 0.019$ per cent.

\begin{figure}
\includegraphics[width=\columnwidth]{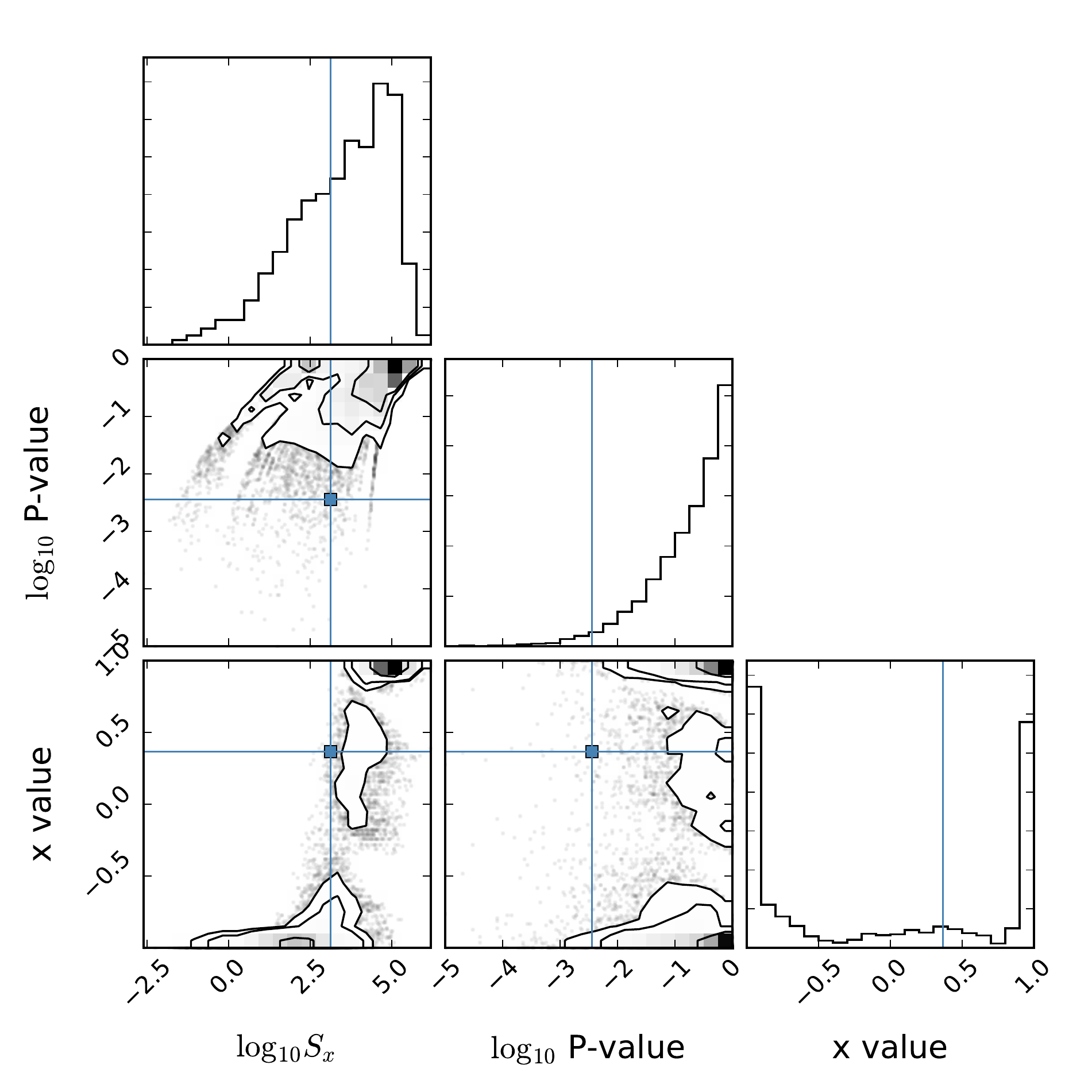}
\caption{Optimized $S_x$, $p(x)$, and $x$ results for $10^4$ minimized $p(x)$ curves, each based on the $S_x$ curves of $10^5$ MC simulated CMBs.  The LEC \pvalue\ of $2.94$ per cent can be seen in the central plot as the small portion of nominal \pvalues\ visible to the left of the vertical line indicating the \texttt{SMICA} value. }
\label{fig:cornerplotNofilter}
\end{figure}

We need to now also account for the fact that we checked the entire range of possible $x$ values in order to find this optimal \pvalue\ (account for the look-elsewhere effect).  We do this by examining the fraction of simulated data sets with x-optimized \pvalues\ less than the \texttt{SMICA} value of $0.36$ per cent, the LEC \pvalue.  For this ensemble, this value is $2.94$ per cent.  Using the small sample variance for the \texttt{SMICA} \pvalue\ described above, this corresponds to  $2.94\%^{+0.14\%}_{-0.07\%}$.

The {\it Planck} team \citep{Pl16XVI} recently implemented a very similar process to this that was meant to account for the look-elsewhere effect caused by the {\it a posteriori} choice of the integration endpoint, and their analysis produced an LEC \pvalue\ of $2.1$ per cent\footnote{The {\it Planck} Collaboration used the opposite sense of \pvalue, such that they calculated the probability of finding values higher than the observed value, rather than lower than the observed value.} for the \texttt{SMICA} map.  However, their analysis was based on only $10^3$ simulations, whereas we used $10^5$.  However, their simulations were much more extensive than ours were.  Their simulations were based on their ``8th Full Focal Plane simulation set'' (FFP8), which includes simulation of instrumental, scanning, and data analysis effects, whereas ours were simply based only on a theoretical power spectrum with some simple Gaussian beam and pixel window smoothing.  Using the a similar procedure as above to estimate sampling uncertainties (with $2.1\% = 21/1000$ and $\sqrt{21}\simeq 4.6$) we find that their result is consistent with ours to within $1.8\sigma$ using their sampling uncertainties.  It is reassuring to note that these two methods gave reasonably consistent results.  

Finally, we combine both methods described here: the ensemble filtering by $C_2$, as well as the $p(x)$ optimization.  In order to reduce computational time, we simply used $10^4$ quadrupole-filtered $S_x$ curves for this.\footnote{The choice of using fewer ($10^4$) simulations is also justified by the larger resultant \pvalue: large numbers of simulations are necessary for high \pvalue\ resolution, which is needed when \pvalues\ are small.  Large numbers of simulations are therefore not needed for larger \pvalues.} The result is shown in Fig.~\ref{fig:cornerplotYesfilter}.  In this ensemble, optimized for \pvalue, the \texttt{SMICA} data produced $x = 0.367$ and $S_x = 1326.5$, similar to the values found without $C_2$ filtering, but a much higher nominal \pvalue\ of $4.50$ per cent, and an even higher LEC \pvalue\ of $22.3$ per cent.

\begin{figure}
\includegraphics[width=\columnwidth]{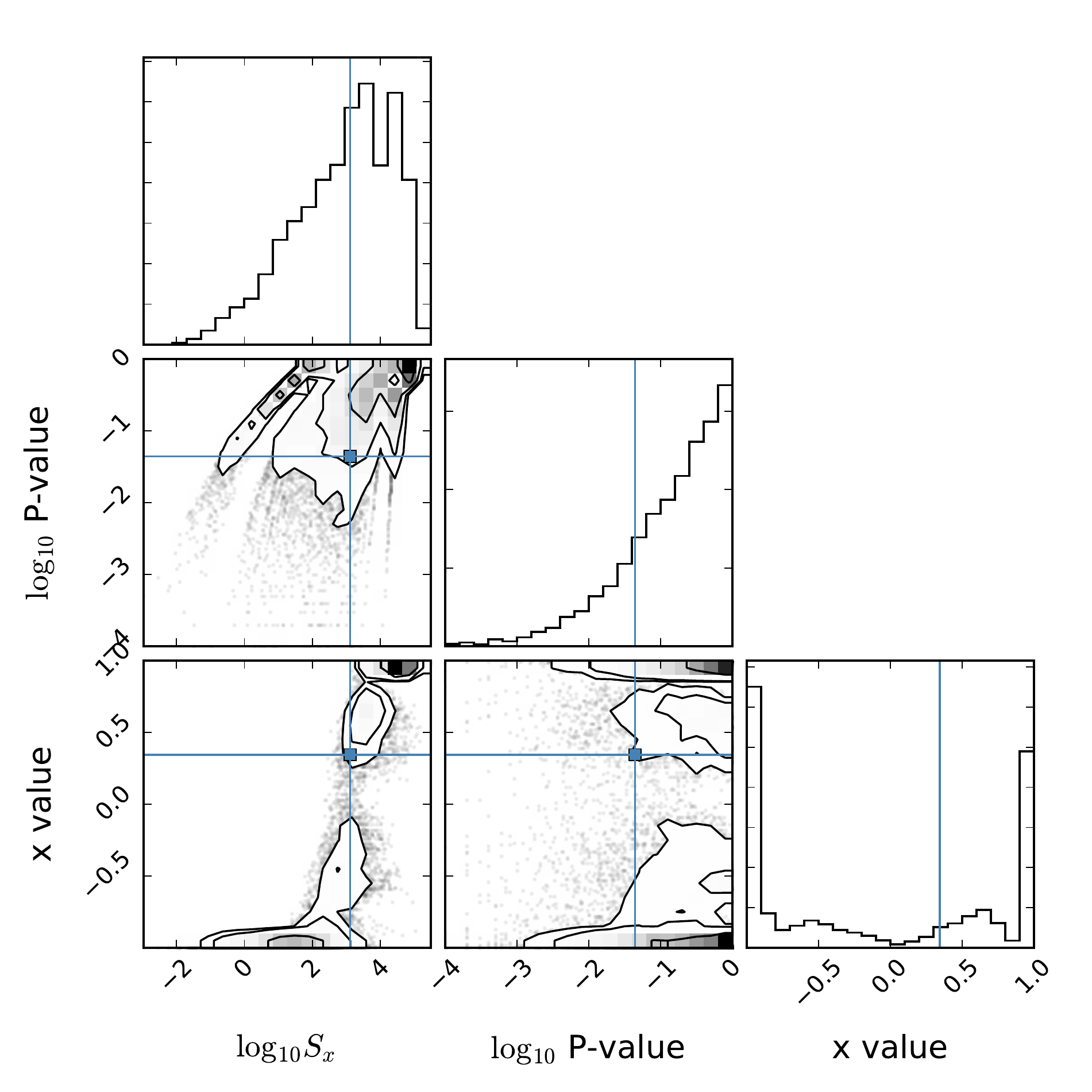}
\caption{Optimized results from $10^4$ MC simulated CMBs, where the simulations have been selected based on having a low quadrupole power after applying the mask. The LEC \pvalue\ of $22$ per cent can be seen in the central plot as the large portion of nominal \pvalues\ visible to the left of the vertical line indicating the \texttt{SMICA} value.} \label{fig:cornerplotYesfilter}
\end{figure}

\section{Summary and Conclusions}\label{summaryandconclusions}

Others have previously noted connections between the low quadrupole and the near-zero correlation function on large angular scales, as quantified in the \Sonehalf\ statistic. Here we quantitatively investigated the relationship between these statistics as expected in the \LCDM\ model. We found that if one conditions on the observed low quadrupole power then the distribution of \Sonehalf\ values in the \LCDM\ model is shifted to much lower values and as a result the observed \Sonehalf\ has a \pvalue\ of 0.08 rather than its unconditioned value of 0.007. We pointed out that departures from \LCDM\ that suppress the quadrupole will thus naturally make the observed \Sonehalf\ less unusual and potentially take it from a \pvalue\ one might call ``anomalous'' to one that is merely ``somewhat unusual'', if that. 

Second, we revisited the exercise in \citet{Pl16XVI}, where \Sonehalf\ was generalized to $S_x$ in order to investigate the impact of the {\it a posteriori} choice of the integration cutoff at $x=\cos\theta = 1/2.$ They found that the process of optimizing the value of $x$ with the observed data leads to a $p$-value that is smaller than what one finds by doing the same process with 21 out of 1000 simulations; i.e., they found a look--elsewhere--corected (LEC) \pvalue\ of 2.1\%. Repeating this exercise with 100 times as many simulations (but without the simulation of as many aspects of the instrument), we found an LEC \pvalue\ of $3$ per cent. Our result is consistent with the $2.1$ per cent value at the $2\sigma$ level given their uncertainties that we estimated from their finite number of samples.  

Third, we combined these two techniques of both conditioning on the observed low quadrupole power and optimizing over $x$ values. This resulted in a quadrupole-filtered LEC \pvalue\ of $22$ per cent.  Our point here is not that a proper calculation of the \pvalue\ has to include conditioning on the low quadrupole, but rather that these are by no means independent phenomena:  One cannot point to both the low quadrupole and low \Sonehalf\ as independent indicators of anomalous behavior on large scales. 

The main thing we offer regarding the LEC \pvalue\ for \Sonehalf\ (quadrupole-filtered or not) is our interpretation of the result. We view the selection of \Sonehalf\ from the one-dimensional space of alternative statistics given by $S_x$ as a proxy for the choice of \Sonehalf\ out of a very high-dimensional space of all possible statistics derivable from \Coftheta\ or $C_\ell$. The value of the proxy is that it allows for calculation of an LEC \pvalue. Since the dimensionality of the alternative space is much lower than that of the space of all possible statistics, it arguably leads to an underestimate of the look-elsewhere correction. In conclusion, we have quantitatively demonstarted that {\it a posteriori} selection provides a viable explanation of the low observed value of \Sonehalf. The correct explanation for low \Sonehalf\ might lie with physics beyond \LCDM\ but there is not yet compelling evidence in favor of such an alternative conclusion. Tests of the so-called ``fluke hypothesis'' \citep{YCSK14, OCKS16} have the potential to alter this situation.

\section*{Acknowledgements}

We thank Brent Follin, Silvia Galli, Marius Millea, Marcio O'Dwyer, and Douglas Scott for useful conversations.

Some of the results in this paper have been derived using the \textsc{healpix} \citep{healpix05} package.

The theoretical power spectrum was computed using the \textsc{class} \citep{BLT11} package.

The cut-sky power spectra were calculated using the \textsc{polspice} (aka \textsc{spice}) \citep{PolSPICE04} package.

The observed CMB data and mask were provided by the {\it Planck} Legacy Archive (\url{http://pla.esac.esa.int/pla/\#home})





\begin{thebibliography}{99}
   
	\bibitem[Addison et. al.(2016)]{Add16} Addison, G. E., Huang, Y., Watts, D. J., Bennett, C. L., Halpern, M., Hinshaw, G., Weiland, J. L., 2016, ApJ, 818, 132
	
	\bibitem[Akrami et. al.(2014)]{AFSEHBG14} Akrami, Y., Fantaye, Y., Shafieloo, A., Eriksen, H. K., Hansen, F. K., Banday, A. J., Górski, K. M., 2014, ApJ, 784L, 42
	
	\bibitem[Aurich \& Lustig(2014)]{AL14} Aurich, R., Lustig, S., 2014, CQGra, 31, 5009
   
   	\bibitem[Bennett et. al.(2003)]{Bea03} Bennett, C. L., Halpern, M., Hinshaw, G., Jarosik, N., Kogut, A., Limon, M., Meyer, S. S., Page, L., Spergel, D. N., Tucker, G. S., Wollack, E., Wright, E. L., Barnes, C., Greason, M. R., Hill, R. S., Komatsu, E., Nolta, M. R., Odegard, N., Peiris, H. V., Verde, L., Weiland, J. L., 2003, ApJS, 148, 1
   
	\bibitem[Bennet et. al.(2011)]{Bea11} Bennett, C. L., Hill, R. S., Hinshaw, G., Larson, D., Smith, K. M., Dunkley, J., Gold, B., Halpern, M., Jarosik, N., Kogut, A., Komatsu, E., Limon, M., Meyer, S. S., Nolta, M. R., Odegard, N., Page, L., Spergel, D. N., Tucker, G. S., Weiland, J. L., Wollack, E., Wright, E. L., 2011, ApJS 192, 17
   
	\bibitem[Bernui et. al.(2006)]{BVWLF06} Bernui, A., Villela, T., Wuensche, C.A., Leonardi, R., Ferreira, I., 2006, A\&A, 454, 409

	\bibitem[Blas, Lesgourges, \& Tram(2011)]{BLT11}  Blas, D., Lesgourgues, J., Tram, T., 2011, JCAP 1107, 034 

	\bibitem[Chon et. al.(2004)]{PolSPICE04} Chon, G., Challinor, A., Prunet, S., Hivon, E., Szapudi, I., 2004, MNRAS, 350, 914

	\bibitem[Contaldi et. al.(2003)]{CPKL03} Contaldi, C. R., Peloso, M., Kofman, L., Linde, S., 2003, JCAP, 07, 002 

	\bibitem[Copi, O'Dwyer, \& Starkman(2016)]{COS16} Copi, C. J., O'Dwyer, M., Starkman, G. D., 2016, MNRAS, 463, 3305
	
	\bibitem[Copi et. al.(2007)]{CHSS07} Copi, C. J., Huterer, D., Schwarz, D. J., Starkman, G. D., 2007, Phys. Rev. D, 75, 023507
	
	\bibitem[Copi et. al.(2009)]{CHSS09} Copi, C. J., Huterer, D., Schwarz, D. J., Starkman, G. D., 2009, MNRAS 399, 295
	
	\bibitem[Copi et. al.(2015a)]{CHSS15} Copi, C. J., Huterer, D., Schwarz, D. J., Starkman, G. D., 2015, MNRAS 451, 2978
	
	\bibitem[Copi et. al.(2015b)]{CHSS15b} Copi, C. J., Huterer, D., Schwarz, D. J., Starkman, G. D., 2015, MNRAS 449, 3458
	
	\bibitem[Cruz, Martínez--González, \& Vielva(2006)]{CTMV06} Cruz, M., Tucci, M., Martínez--González, E., Vielva, P., 2006, MNRAS, 369, 57
	
	\bibitem[Cruz, Martínez--González, \& Vielva(2007)]{CCMVJ07} Cruz, M., Cayón, L., Martínez--González, E., Vielva, P., Jin, J., 2007, ApJ, 655, 11
	
	\bibitem[Cruz, Martínez--González, \& Vielva(2010)]{CMV10} Cruz, M., Martínez--González, E., Vielva, P., 2010, ASSP, 14, 275
	
	\bibitem[de Oliveira--Costa et. al.(2004)]{dTZH04} de Oliveira--Costa, A., Tegmark, M., Zaldarriaga, M., Hamilton, A., 2004, Phys. Rev. D., 69, 063516
	
	\bibitem[Efstathiou(2003)]{E03} Efstathiou, G., 2003, MNRAS, 346, L26
	
	\bibitem[Efstathiou, Ma, \& Hanson(2010)]{EMH10} Efstathiou, G., Ma, Y., Hanson, D., 2010, MNRAS, 407, 2530
	
	\bibitem[Eriksen et. al.(2004)]{EHBGL04} Eriksen, H. K., Hansen, F. K., Banday, A. J., Górski, K. M., Lilje, P. B., 2004, ApJ, 605, 14
	
	\bibitem[Eriksen et. al.(2007)]{EBGHL07} Eriksen, H. K., Banday, A. J., Górski, K. M., Hansen, F. K., Lilje, P. B., 2007, ApJ, 660, L81
	
	\bibitem[Fixsen(2009)]{F09} Fixsen, D.J., 2009, ApJ, 707, 916

	\bibitem[Foreman--Mackey(2016)]{F16} Foreman--Mackey, D., 2016, JOSS, 24
	
	\bibitem[Frolop \& Scott(2016)]{FS16} Frolop, A., Scott, D., 2016, arXiv:1603.09703

	\bibitem[Górski et. al.(2005)]{healpix05} Górski, K.M., Hivon, E., Banday, A.J., Wandelt, B.D., Hansen, F.K., Reinecke, M., Bartelmann, M., 2005, Ap.J., 622, 759

	\bibitem[Gruppuso(2014)]{G14} Gruppuso, A., 2014, MNRAS 437, 2076
	
	\bibitem[Gruppuso \& Sagnotti(2014)]{GS14} Gruppuso, A., Sagnotti, A., 2014, IJMPD, 24, 1544008-469 
	

	\bibitem[Hansen et. al.(2004)]{HCMV04} Hansen, F. K., Cabella, P., Marinucci, D., Vittorio, N., 2004, ApJL, 607, L67

	\bibitem[Hansen et. al.(2009)]{HBGEL09} 	Hansen, F. K., Banday, A. J., Górski, K. M., Eriksen, H. K., Lilje, P. B., 2009, ApJ, 704, 1448

	\bibitem[Hajian(2007)]{H07} Hajian, A., 2007, preprint (astro-ph/0702723)

	\bibitem[Hinshaw et. al.(1996)]{H96} Hinshaw, G., Branday, A. J., Bennett, C. L., Gorski, K. M., Kogut, A., Lineweaver, C. H., Smoot, G. F., Wright, E. L., 1996, ApJ, 464, L25
	
	\bibitem[Hoftuft et. al.(2009)]{HEBGHL09} Hoftuft, J., Eriksen, H. K., Banday, A. J., Górski, K. M., Hansen, F. K.; Lilje, P. B., 2009, ApJ, 699, 985
	
	\bibitem[Kim \& Naselsky(2010a)]{KN10a} Kim, J., Naselsky, P., 2010, ApJ, 714, 265
	
	\bibitem[Kim \& Naselsky(2010b)]{KN10b} Kim, J., Naselsky, P., 2010, Phys. Rev. D., 82, 0603002
	
	\bibitem[Kim \& Naselsky(2011)]{KN11} Kim, J., Naselsky, P., 2011, ApJ, 739, 79
	
	\bibitem[Land \& Magueijo(2005a)]{LM05a} Land, K., and Magueijo, J., 2005, Phys. Rev. D., 72, 101302
	
	\bibitem[Land \& Magueijo(2005b)]{LM05b} Land, K., and Magueijo, J., 2005, Phys. Rev. D., 95, 071301
	
	\bibitem[Liu et. al.(2014)]{LGP14}Liu, Z., Guo, Z., Piao, Y., 2014, EPJC, 74, 3006L

	\bibitem[Lello et. al.(2014)]{LBH14} Lello, L., Boyanovsky, D., Holman, R., 2014, Phys. Rev. D., 89, 063533


	\bibitem[Niarchou \& Jaffe(2006)]{NJ06} Niarchou, A., and Jaffe, A. H., 2006, AIPC, 848, 774
	
	\bibitem[O'Dwyer et. al.(2016)]{OCKS16} O'Dwyer, M., Copi, C. J., Knox, L., Starkman, G. D., 2016, arXiv:1608.02234
	
	\bibitem[Planck XV(2014)]{Pl14XV} Planck Collaboration XV, 2014, A\&A, 571, A15
	
	\bibitem[Planck XVII(2014)]{Pl14XVII} Planck Collaboration XVII, 2014, A\&A, 571, A17
	
	\bibitem[Planck XXIII(2014)]{Pl14XXIII} Planck Collaboration XXIII, 2014, A\&A, 571, A23

	\bibitem[Planck XXVII(2014)]{Pl14XXVII} Planck Collaboration XXVII, 2014, A\&A, 571, A27	

	\bibitem[Planck XI(2016)]{Pl16XI} Planck Collaboration XI, 2016, A\&A, 594, A11

	\bibitem[Planck XIII(2016)]{Pl16XIII} Planck Collaboration XIII, 2016, A\&A, 594, A13

	\bibitem[Planck XV(2016)]{Pl16XV} Planck Collaboration XV, 2016, A\&A, 594, A15

	\bibitem[Planck XVI(2016)]{Pl16XVI} Planck Collaboration XVI, 2016, A\&A, 594, A16
	
	\bibitem[Planck LI(2016)]{Pl16LI} Planck Collaboration LI, 2016, arXiv: 1608.02487
	
	\bibitem[Schwarz et. al.(2004)]{SSHC04} Schwarz, D. J., Starkman, G. D., Huterer, D., Copi, C. J., 2004, Phys. Rev. Let., 93, 221301
	
	\bibitem[Schwarz et. al.(2015)]{SCHS15} Schwarz, D.J., Copi, C.J., Huterer, D., Starkman, G.,  2015, arXiv: 1510.07929v1
	
	\bibitem[Spergel et. al.(2003)]{Sp03} Spergel, D. N., Verde, L., Peiris, H. V., Komatsu, E., Nolta, M. R., Bennett, C. L., Halpern, M., Hinshaw, G., Jarosik, N., Kogut, A., Limon, M., Meyer, S. S., Page, L., Tucker, G. S., Weiland, J. L., Wollack, E., Wright, E. L., 2003, ApJS, 148, 175
	
	\bibitem[Stevens, Silk, \& Scott(1993)]{SSS93} Stevens, D., Scott, D., Silk, J., 1993, Phys Rev Lett, 71, 20
	
	\bibitem[Yoho et. al.(2014)]{YCSK14} Yoho, A., Copi, C. J., Starkman, G. D., Kosowsky, A., 2014, MNRAS, 442, 2392
	
	\bibitem[Zhang \& Huterer(2010)]{ZH10} Zhang, R., Huterer, D., 2010, APh, 33, 69

\end{thebibliography}




\appendix

\section{Comparison to other Results}

In order to help verify our methods, we approximately repeated one of the \Sonehalf\ calculations done by \cite{CHSS15}, attempting to use the same methods and data that they had used, in order to obtain the same results.  One of the data sets which they used included the {\it Planck} PR1 \texttt{SMICA} map and the so-called \texttt{U74} mask, for which \cite{CHSS15} found \Sonehalf\ $= 1577.7~{\rm \mu K}^4$,  and using an ensemble of $10^6$ simulations, a \pvalue\ of $0.191$ per cent.  We obtained the publicly available PR1 \texttt{SMICA} map and the \texttt{U73} mask, but did not attempt to reproduce the \texttt{U74} map, which was designed as an approximation to \texttt{U73}, since it was not publicly available at the time of the previous analysis.  We also modified our mask and map degradation procedures slightly to match what \cite{CHSS15} did in their analysis: they used the \textsc{healpix} \texttt{ud\_grade} function, rather than the harmonic space window-weighting method, as well as a mask threshold of 0.8, rather than our 0.9.

We nearly reproduced their result, finding \Sonehalf\ $= 1527.5~{\rm \mu K}^4$, and with an ensemble of $10^5$ simulations, found a \pvalue\ of $0.247$ per cent.  To estimate the sampling uncertainty, as in section \ref{noAPosteriori}, we suppose that our measured \Sonehalf\ \pvalue\ was drawn from a Poissonian distribution and calculate the standard deviation as $\sqrt{247}/10^5 \simeq .016$ per cent.  Thus the difference between our p-value and that of \cite{CHSS15} is about $3.6\sigma$.  This difference in \pvalues\ is likely due to the difference in the masks that we used (\texttt{U73} vs. \texttt{U74}).


\bsp	
\label{lastpage}
\end{document}